\documentclass{article}

\usepackage[english]{babel}

\usepackage[letterpaper,top=2cm,bottom=2cm,left=3cm,right=3cm,marginparwidth=1.75cm]{geometry}

\usepackage{amsmath}
\usepackage{graphicx}
\usepackage{subcaption}
\usepackage{amssymb}
\usepackage[colorlinks=true, allcolors=blue]{hyperref}

\title{Enhancing Software Effort Estimation through Reinforcement Learning-based Project Management-Oriented Feature Selection}
\author{Haoyang Chen}

\begin{document}
\maketitle

\begin{abstract}
\noindent \textbf{Purpose} - The study aims to investigate the application of the data element market in software project management, focusing on improving effort estimation by addressing challenges faced by traditional methods.

\noindent \textbf{Design/methodology/approach} - This study proposes a solution based on feature selection, utilizing the data element market and reinforcement learning-based algorithms to enhance the accuracy of software effort estimation. It explores the application of the MARLFS algorithm, customizing improvements to the algorithm and reward function.

\noindent \textbf{Findings} - This study demonstrates that the proposed approach achieves more precise estimation compared to traditional methods, leveraging feature selection to guide project management in software development.

\noindent \textbf{Originality/value} - This study contributes to the field by offering a novel approach that combines the data element market, machine learning, and feature selection to improve software effort estimation, addressing limitations of traditional methods and providing insights for future research in project management.
\end{abstract}

\section{Introduction}

With the rapid development of information technology and the rise of data-driven decision-making, the data element market, as an emerging market, is gradually attracting attention. In this market, data is considered a valuable resource that can be used to train machine learning models, support artificial intelligence solutions, and conduct various data-driven business activities. The emergence of the data element market not only provides enterprises and research institutions with a way to access high-quality data but also brings new opportunities and challenges to software project management.

Software project management refers to the process of effectively planning, organizing, directing, and controlling to complete software development projects. In traditional software project management, effort estimation is a critical step that determines the project's schedule, cost, and resource allocation. However, in reality, effort estimation often faces many challenges, such as the lack of accurate historical data and the uncertainty caused by subjective assessments\cite{qikan-zongshu}.

In software project management, traditional methods for effort estimation generally include estimation approaches such as PERT networks\cite{pert1,pert2}, COCOMO model\cite{cocomo1}, expert judgment-based methods\cite{zhuanjia1}, function point-based methods\cite{gongnengdian1,gongnengdian2}, and UCP (Use Case Points)\cite{ucp1}. With the development of big data technology and the rise of machine learning, an increasing number of studies suggest that integrating machine learning into effort prediction can achieve better results. For example, Akshay Jadhav et al. concluded that machine learning improves the performance of Software Development Effort Estimation based on experiments and comparative metrics\cite{AkshayMLEE}. A. B. Nassif et al. employed a cascaded correlation neural network approach to predict software effort from use case diagrams\cite{NassifMLEE}. Fatemeh Zare et al. proposed a software effort (person-month) estimation model based on a three-layer Bayesian network and COCOMO, considering 15 components and software scale\cite{FatemehMLEE}. Safae Laqrichi et al. introduced a method to incorporate uncertainty into neural network-based effort estimation models\cite{SafaeMLEE}. Ali Bou Nassif et al. developed a Multi-Layer Perceptron (MLP) neural network model for predicting software effort based on software scale and team productivity, demonstrating the superiority of their proposed method over the original UCP model\cite{AliMLEE}.

Based on machine learning, software effort estimation generally offers greater adaptability and more accurate forecasts compared to traditional methods. However, it also comes with some common drawbacks, such as the need for large amounts of historical data for training and the difficulty in interpreting its prediction results. Through feature selection, one can both proactively identify which features are crucial in influencing software effort, providing guidance for project management, and leverage the advantages of machine learning to make predictions superior to traditional methods.

As a platform offering diverse data resources, the Data Element Market provides abundant material for feature selection. Through datasets available in the market, project management teams can access a wealth of project historical data, including task assignments across different project phases, completion times, issues, and solutions. These data serve as valuable assets for conducting feature selection, revealing key factors influencing effort of software development.

The following section explores the enhancement of effort estimation in software project management through feature selection, using MARLFS (Multi-Agent Reinforcement Learning Feature Selection) as an example. MARLFS is a multi-agent reinforcement learning feature selection algorithm that demonstrates excellent performance across multiple performance metrics, making it an important tool in the Data Element Market for automation and performance\cite{MARLFS}.

In current research, the application of reinforcement learning (RL) in feature selection has attracted widespread attention. By learning through interaction with the environment, RL can effectively address issues such as high dimensionality, non-linearity, and incompleteness in feature selection, making it promising for application in project management. Furthermore, existing studies have demonstrated the feasibility of RL in feature selection and have achieved significant performance improvements in multiple domains. For example, Seyed Mehdi Hazrati Fard et al. treated feature selection as a reinforcement learning problem and used temporal difference methods to find the optimal feature subset\cite{SeyedRL}. Lu Liu et al. proposed a two-stage feature selection method called CorrDQN-FS, which combines Pearson correlation coefficient and deep reinforcement learning techniques to improve the accuracy of energy consumption prediction models\cite{luliuRL}. Zhengpeng Hu et al. introduced a reinforcement learning-based Comprehensive Learning Grey Wolf Optimizer (RLCGWO) to solve feature selection problems. By designing comprehensive learning operators and RL-based policy adjustment techniques, combined with chaos-based learning strategies, they significantly improved convergence speed and accuracy\cite{ZhengpengRL}. Seo-Hee Kim et al. developed a fair feature selection algorithm aimed at proportionally allocating rewards based on agents' contributions to optimize feature extraction and improve model performance\cite{SeoRL}. Mohsen Paniri et al. proposed a novel multi-label feature selection method based on ant colony optimization. By optimizing the ACO heuristic function using heuristic learning methods, they demonstrated superior classification performance over competitive methods on nine benchmark datasets\cite{PANIRI2021100892}.

Indeed, while MARLFS excels in classification tasks, it is not directly applicable to regression tasks such as software effort estimation. Therefore, utilizing MARLFS for feature selection in software effort estimation requires some additional customization and adaptation.

The first step involves enhancing the MARLFS algorithm to cater to the requirements of regression tasks. Initially, MARLFS embeds classifiers internally, so it needs to be modified to embed regressors instead, suitable for regression tasks. This step is crucial as the algorithm's performance of regressors directly impacts the effectiveness and robustness of feature selection.

Next, it's necessary to tailor a reward function suitable for software effort prediction. The MARLFS paper lacks details regarding the reward function, and for regression tasks, the design of the reward function is particularly crucial. A suitable reward function should accurately measure the model's bias, thereby guiding the feature selection algorithm to select the optimal feature subset.

By overcoming these challenges, the effective application of advanced feature selection methods in software project management can be realized, aiming to enhance the accuracy and reliability of effort estimation. This not only aids in improving project management efficiency but also provides a more scientific basis for project decision-making.

By introducing feature selection into the data element market, software project management teams expect to achieve the following advantages: (1) Accurate Effort Estimation: By filtering and selecting the most relevant features, the complexity of the model can be reduced, leading to more accurate effort estimations and enhancing the accuracy of project management. (2) Risk Reduction: Feature selection helps identify and retain the most valuable information, assisting project teams in better understanding the critical factors of the project, thereby reducing estimation uncertainty and project risks. (3) Improved Resource Utilization: Through refined feature selection, project resources can be allocated more effectively, ensuring that key tasks receive sufficient attention and thereby enhancing overall resource utilization.

The main contributions of this paper are as follows: (1) Proposing a reward function suitable for regression tasks based on reinforcement learning. (2) Introducing MARLFS for regression tasks. (3) Proposing the utilization of reinforcement learning-based feature selection to address software effort estimation issues within the context of the data element market. (4) Suggesting the use of feature selection results to guide software project management.

\section{Methods }

\subsection{Expert Judgment }

Traditional methods of effort estimation typically rely on experience and historical data, such as function point-based methods\cite{gongnengdian2} and the COCOMO model\cite{cocomo1}. While these methods can provide estimates of effort to some extent, they are often limited by data constraints, model assumptions, and the level of understanding of project characteristics.

Among them, expert judgment is a common traditional method that relies on the subjective judgment and experience of domain experts\cite{zhuanjia2}.

The expert judgment method typically involves the following steps:

\begin{enumerate}
    \item Selecting Expert Team Members: Choose domain experts with extensive experience and professional knowledge to form an expert team.
    \item Collecting Project Information: Gather project-related information, such as project size, complexity, technical requirements, etc.
    \item Expert Assessment: Experts assess the workload of the project based on their experience and judgment, and provide estimation values.
    \item Summarizing and Analyzing: Summarize the assessments of the experts, analyze and discuss them, and ultimately determine the effort estimation results.
\end{enumerate}
The advantages of expert judgment lie in its speed and flexibility, making it suitable for small projects or situations where sufficient data support is lacking. However, expert judgment also has some disadvantages, such as being heavily influenced by individual experience and subjectivity, prone to estimation biases, and difficult to maintain consistency in large or complex projects.

\subsection{Random Forest }
\begin{figure}
\centering
\includegraphics[width=0.8\linewidth]{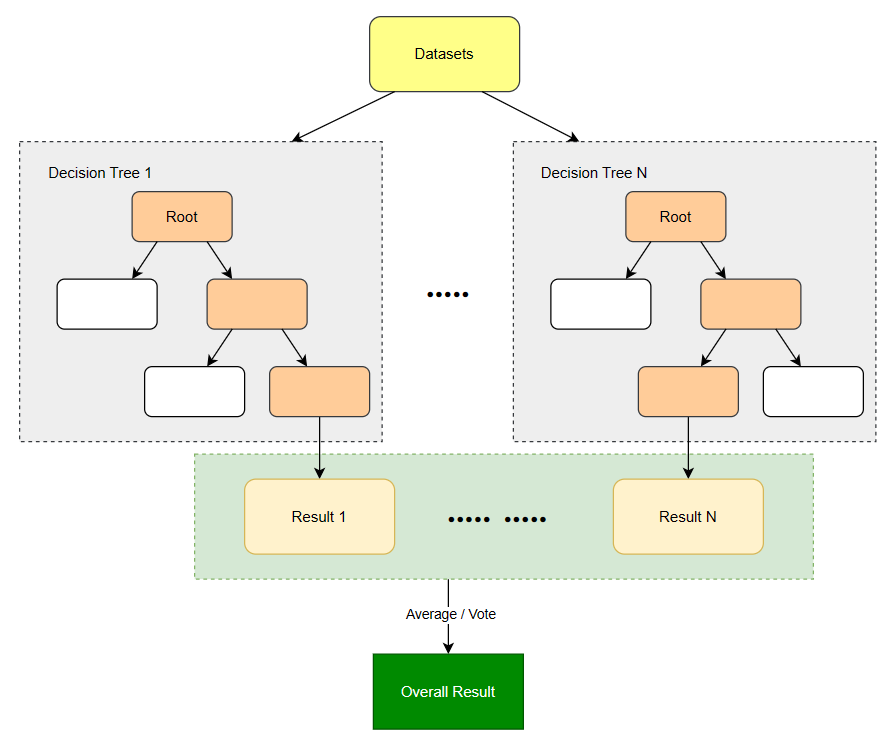}
\caption{\label{fig:rf}Random Forest Algorithm}
\end{figure}

Random Forest is an ensemble learning algorithm (Bagging) commonly used for classification and regression tasks\cite{breiman2001rf}. As illustrated  in the figure \ref{fig:rf} , it is based on decision tree construction, training multiple decision trees, and combining their prediction results to improve model performance and generalization ability. The characteristics of training decision trees in Random Forest are as follows: (1) Random Sampling: Randomly sample a certain number of samples from the original dataset, with replacement, to train each decision tree. This bootstrap sampling technique ensures the diversity of training sets for each decision tree. (2) Random Feature Selection: During the construction of each decision tree, randomly select a certain number of features as candidate features. This increases the diversity of each decision tree and improves the model's generalization ability. (3) Independent Training: Each decision tree is trained independently, without any dependency between them. (4) Voting or Averaging: For classification problems, Random Forest uses a voting mechanism where each decision tree votes to give the final classification result. For regression problems, Random Forest uses the average prediction result. 

In scikit-learn, the RandomForestRegressor employs the CART (Classification and Regression Trees) algorithm by default for decision trees. CART is a tree algorithm that can be used for both classification and regression tasks\cite{cartrf}. It differs from ID3 and C4.5 in that it constructs decision trees using the Gini impurity criterion. The Gini impurity is a measure of the impurity of a dataset, indicating the probability of selecting two samples from the dataset with different class labels. It is defined as follows: 
\begin{equation}
    \text{Gini}(D) = 1 - \sum_{i=1}^{k} (p_i)^2
\end{equation}

$D$ represents the dataset, $k$ is the number of categories in the dataset, and $p_i$ is the proportion of samples belonging to category $i$ in the dataset.
The Gini impurity ranges from 0 to 1, where a value closer to 0 indicates higher purity of the dataset, and a value closer to 1 indicates higher impurity. When the dataset contains only one category, the Gini impurity reaches its minimum value of 0, indicating complete purity of the dataset. When the number of samples in each category is equal, the Gini impurity reaches its maximum value of 0.5, indicating the dataset is most impure.

\subsection{Feature Selection }

\subsubsection{Filter}
The Filter feature selection method is a technique for evaluating the importance of features based on their intrinsic statistical properties, independent of any specific machine learning algorithm\cite{filter1}. This approach assesses and ranks features by computing their correlation or other statistical metrics with the target variable. Common metrics used in Filter feature selection methods include correlation coefficient, analysis of variance (ANOVA), information gain, Chi-Square test, and mutual information\cite{filter2}.
The advantages of Filter feature selection methods include simplicity in computation, speed, and independence from specific machine learning algorithms. However, they also have some limitations, such as the inability to capture complex relationships between features and the potential neglect of interactions among features.

\subsubsection{Wrapper }
The Wrapper method is a feature selection technique based on machine learning models, which selects the optimal feature subset by training a predictive model and evaluating the performance based on model performance\cite{wrapper}. When employing the Wrapper method to search for the best feature subset, it's common to integrate cross-validation to assess the performance of each feature subset and select the combination of features with the best performance.

Assuming we have a dataset $\mathbf{X}$ consisting of $n$ samples and $m$ features.

1. Define Feature Subset:
    
    We begin by defining a feature subset. For instance, with $m$ features, we represent the feature subset as  $\mathbf{S} = \{s_1, s_2, ..., s_k\}$, where  $k \leq m$, and $s_i$ indicates whether the $i$-th feature is included in the subset. It's represented as 1 if included, and 0 otherwise. For example, with 5 features, a possible feature subset could be $\mathbf{S} = \{1, 0, 1, 1, 0\}$.
    
2. Define Evaluation Function:
    
    We need to define an evaluation function $E(\mathbf{S})$ to assess the performance of the feature subset $\mathbf{S}$. This evaluation function typically depends on our problem and model selection. For instance, in effort estimation regression problems, it could be Mean Squared Error (MSE).
    
3. Search for Optimal Feature Subset:
    
    Our objective is to find the feature subset $\mathbf{S}$ that minimizes (or maximizes, depending on the problem type) the evaluation function $E(\mathbf{S})$. This is a combinatorial optimization problem, and various methods such as exhaustive search, greedy search, or genetic algorithms can be employed to find the optimal feature subset.
    
4. Cross-Validation:
    
    During the evaluation of each feature subset, cross-validation is commonly used to estimate the model's performance on unseen data. This helps prevent overfitting and provides a better assessment of the model's generalization ability.
    
5. Selection of Optimal Feature Subset:
    
    After the search and cross-validation, we obtain multiple feature subsets along with their corresponding evaluation function values. The feature subset with the best performance is chosen as the final feature combination.

In the end, we select the feature subset $\mathbf{S}^*$ with the minimum cross-validation error $E_{CV}(\mathbf{S})$ as the optimal feature combination.

Wrapper feature selection methods comprehensively consider interactions among features, but they come with high computational costs and are prone to overfitting, making them unsuitable for large-scale feature spaces.

\subsubsection{MARLFS }

MARLFS (Multi-Agent Reinforcement Learning Feature Selection) is a feature selection method based on multi-agent reinforcement learning. Its aim is to identify the most valuable feature subset by leveraging cooperation and competition among agents, thereby enhancing the performance of machine learning models\cite{MARLFS}. This method is particularly suitable for handling high-dimensional data and complex relationships among features. The basic steps of the MARLFS feature selection method are as follows:

1. Initializing Multiple Agents: In MARLFS, each agent corresponds to a feature. The task of each agent is to learn when to select or deselect its corresponding feature through interaction with the environment.

2. Defining the Reward Function: The reward function serves as the criterion for evaluating the goodness of agent behavior, typically directly related to the performance of the feature subset (such as classification accuracy or regression error). The goal of the agents is to maximize their cumulative reward.

3. Environment State Representation: In MARLFS, the environment state is composed of the selected features. Since the number of selected features can vary, leading to changes in the length of the state vector, a method needs to be defined to aggregate different numbers of features into a fixed length.

4. Agent Exploration and Exploitation: Agents employ DQN (Deep-Q Network) to explore (trying out new feature combinations) and exploit (using known best feature combinations) in learning how to select features. This process involves strategies from reinforcement learning such as $\varepsilon$-greedy policy or Upper Confidence Bound (UCB) policy, as well as the design of the Replay Buffer. During the training process of DQN, the Bellman equation is used to compute the target Q-value. Specifically, for a transition $(s, a, r, s')$, the target Q-value is computed based on the network's current estimate of Q-values, following the Bellman equation:
\begin{equation}
y = r + \gamma \max_{a'}Q(s', a'; \theta^-)
\end{equation}

In this context,  $\theta^-$ represents the parameters of the target network. During the training process, DQN updates the network parameters by minimizing the difference between the target Q-values and the estimated Q-values, typically using mean squared error as the loss function. Through this approach, DQN learns a policy that enables each agent to choose the optimal action given the state representation provided at step 3, maximizing the long-term reward.

5. Collaboration and Competition Mechanism: In multi-agent systems, agents can share knowledge through collaboration, for example, by exchanging feature information they consider important. At the same time, competition among agents drives them to seek unique and effective feature subsets to achieve higher rewards.

6. Feature Subset Evaluation and Selection: After each round of exploration and exploitation, each agent selects or deselects corresponding features to form feature subsets. These feature subsets are evaluated based on their performance on the training model, such as cross-validation scores. Ultimately, the best-performing feature subset is chosen as the final result.

7. Model Training and Validation: Machine learning models are trained using the selected feature subsets and validated on an independent test set to ensure the effectiveness of the feature selection process and the generalization ability of the model.

\begin{figure}
\centering
\includegraphics[width=0.8\linewidth]{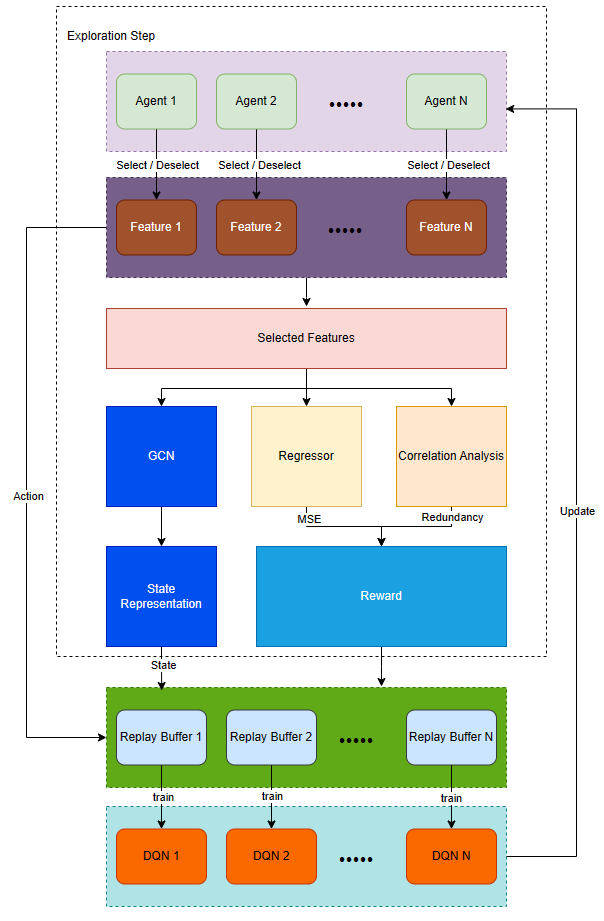}
\caption{\label{fig:MARLFS}MARLFS Algorithm}
\end{figure}

The MARLFS design achieves a highly adaptive and efficient feature selection method, which has consistently outperformed other feature selection approaches in previous experiments. Therefore, in this study, we replicated this method and modified it to suit regression tasks such as software effort estimation, conducting feature selection experiments on the SEERA dataset. Specifically, the structure of the implemented MARLFS is illustrated in the figure \ref{fig:MARLFS}.

\subsection{Study of Reward Function  }

In reinforcement learning, the reward function plays a crucial role in guiding the learning of the agent. It serves as a mechanism for evaluating the agent's behavior by providing positive or negative rewards to encourage the agent to achieve specific goals. The design of the reward function determines the behavior of the agent during the learning process. Since the original MARLFS algorithm was designed for classification tasks and lacks details on the design of the reward function, we need to redesign a reward function for our regression task of software effort estimation. A good reward function should motivate the agent to take the correct actions, while the opposite could lead to the learning of undesirable strategies. Therefore, the design of the reward function needs to avoid unnecessary complexity while ensuring it effectively guides the agent to learn the desired behavior. In our task, we plan to incorporate metrics such as mean squared error for regression algorithm performance measurement and the Pearson correlation coefficient between features as a measure of redundancy into the reward function.

The Mean Squared Error (MSE) is a commonly used metric for evaluating the performance of regression models. It calculates the average of the squared differences between the predicted values of the model and the actual observed values. The MSE value is a non-negative real number, ranging from $[0, +\infty)$. A lower MSE value indicates higher accuracy of the model's predictions, while a larger value indicates greater prediction error. Specifically, if there are $n$ data points, and the model's predicted value for the $i$-th observation is $\hat{y}_i$, while the actual observed value is $y_i$, then the MSE is defined as:

\begin{equation}
    \text{MSE} = \frac{1}{n} \sum_{i=1}^{n} (\hat{y}_i - y_i)^2 \label{eq:mse}
\end{equation}

The Pearson correlation coefficient is a statistical measure of the linear relationship between two variables, with values ranging from $[-1, 1]$. A correlation coefficient of $r=1$ indicates a perfect positive linear relationship between the two variables, while $r=-1$ indicates a perfect negative linear relationship. A correlation coefficient of $r=0$ suggests no linear relationship between the variables. The formula for calculating the Pearson correlation coefficient is as follows:
\begin{equation}
    r = \frac{\sum_{i=1}^{n} (x_i - \bar{x})(y_i - \bar{y})}{\sqrt{\sum_{i=1}^{n} (x_i - \bar{x})^2} \sqrt{\sum_{i=1}^{n} (y_i - \bar{y})^2}}
\end{equation}
Note that $x_i$ and $y_i$ represent the observed values of the two variables, while $\bar{x}$ and $\bar{y}$ denote their respective means.

Redundancy in statistics and data analysis typically refers to the repetition of information within a dataset. When considering features, if one variable can be predicted or explained by another variable, then there is redundancy between these two features. When the Pearson correlation coefficient between two features approaches 1 or -1, it indicates a strong linear relationship between them, implying that one variable can be well predicted by another, leading to data redundancy. Therefore, we use the absolute value of the Pearson correlation coefficient, $|r| \in [0, 1]$, to represent redundancy. This serves as a negative indicator for the reward function. For the redundancy of selected feature subsets, we distribute the total redundancy among individual features, yielding the average redundancy denoted as $\bar{r}$. By incorporating redundancy as a term in the reward function, we aim to mitigate the occurrence of multicollinearity among features.

Incorporating the mean squared error (MSE) alongside an existing redundancy metric with known range values, we require a transformation function to map MSE$\in [0, +\infty]$ onto the interval $[0,1]$. Furthermore, the transformed function should serve as a positive indicator for the reward function, meaning that as MSE approaches 0, the transformed function and consequently the reward function should increase. Considering two candidate functions based on their graphical analysis: 

(1)  $2(1-sigmoid(x))$   (2) $e^{-x}$.

The mathematical expression of the Sigmoid function is as follows:
\begin{equation}
    \sigma(x) = \frac{1}{1 + e^{-x}}
\end{equation}

\begin{figure}
    \centering
    \begin{subfigure}[b]{0.4\textwidth}
        \centering
        \includegraphics[width=\textwidth]{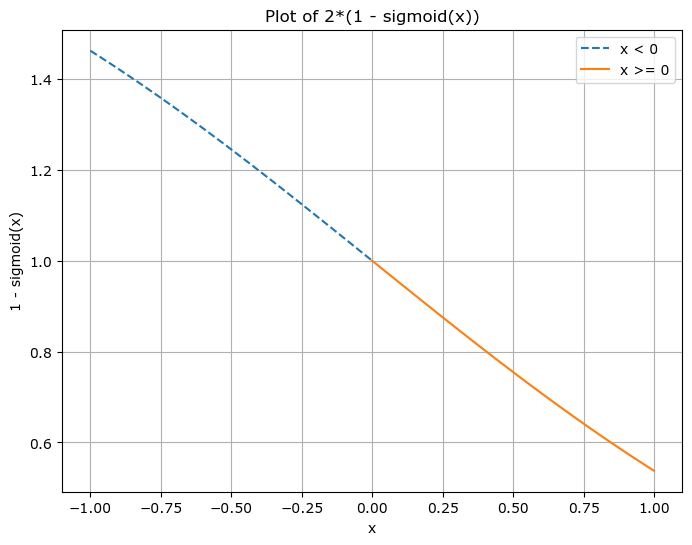}
        \caption{Plot of 2*(1-sigmoid(x)) }
        \label{fig:plot_2}
    \end{subfigure}
    \hfill
    \begin{subfigure}[b]{0.4\textwidth}
        \centering
        \includegraphics[width=\textwidth]{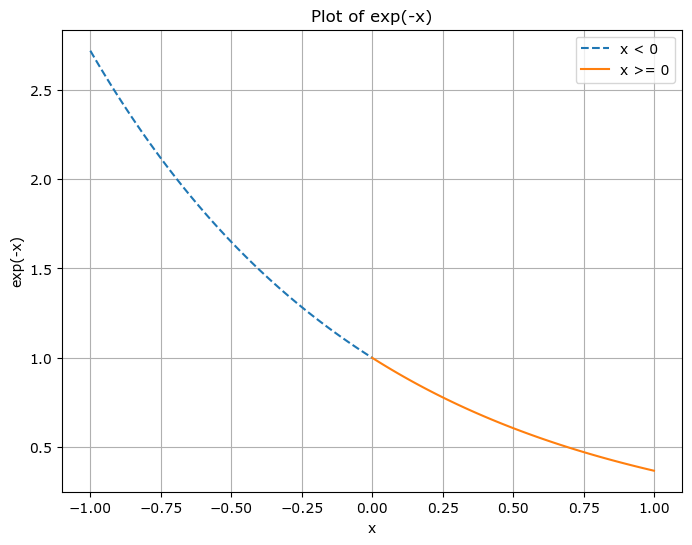}
        \caption{Plot of exp(-x) }
        \label{fig:plot_exp}
    \end{subfigure}
    \caption{Comparison of  proposed functions.}
    \label{fig:plot_functions}
\end{figure}

\begin{figure}
    \centering
    \begin{subfigure}[b]{0.4\textwidth}
        \centering
        \includegraphics[width=\textwidth]{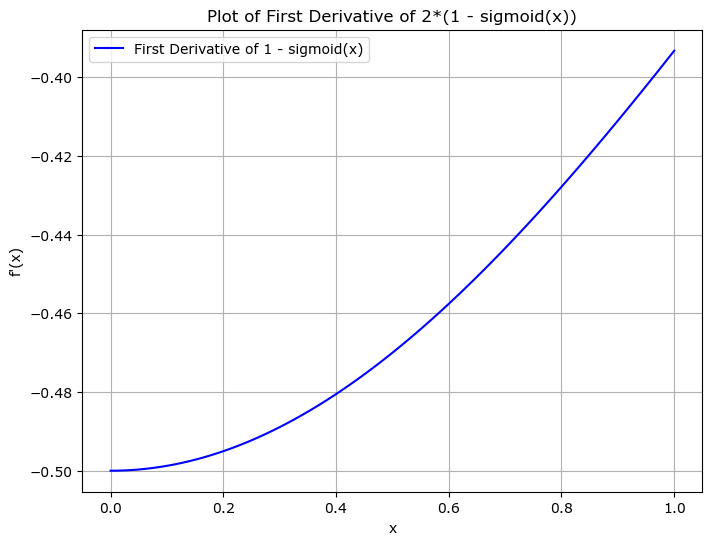}
        \caption{Plot of derivatives of 2*(1-sigmoid(x)) }
        \label{fig:plot_d_2}
    \end{subfigure}
    \hfill
    \begin{subfigure}[b]{0.4\textwidth}
        \centering
        \includegraphics[width=\textwidth]{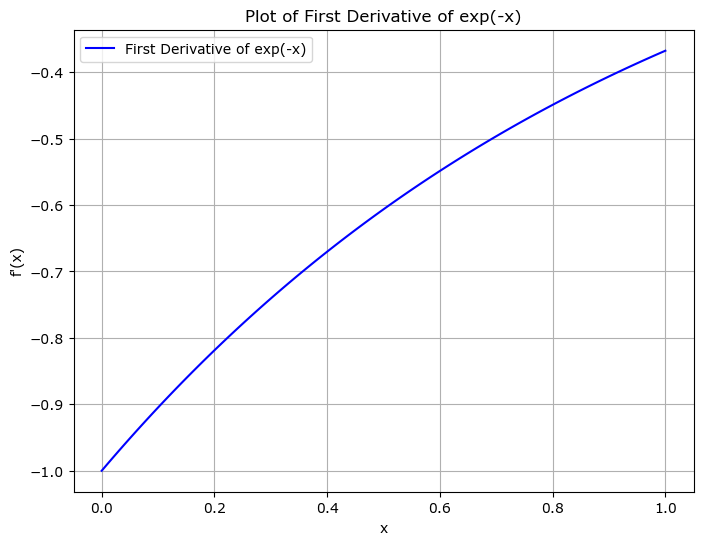}
        \caption{Plot of derivatives of exp(-x) }
        \label{fig:plot_d_exp}
    \end{subfigure}
    \caption{Comparison of the derivatives of proposed functions.}
    \label{fig:plot_derivatives}
\end{figure}

The graphs of the two functions are shown in the figure \ref{fig:plot_functions} , both capable of mapping MSE to the interval $[0,1]$. However, when considering the first derivative, as figure \ref{fig:plot_derivatives}, compared to the function $2(1-sigmoid(x))$ , the exponential function $e^{-x}$ approaches 0, causing the gradient to increase. When applied to a reward function, this encourages MSE to approach 0. Therefore, for the transformation function, we use the exponential function. Additionally, since we need to initially reduce the input MSE to a certain extent, we introduce a decay factor k, resulting in $e^{-kx}$.

Building upon this, we integrate the obtained two metrics, based on the positivity or negativity of the reward, into one equation. We then introduce weight hyperparameters $\alpha$ and $\beta$ corresponding to the two metrics, yielding the expression for the reward function:
\begin{equation}
    reward = \alpha e^{-kMSE}-\beta \bar{r}
\end{equation}

\section{Experimental Process }

\subsection{Dataset Preprocessing }

In this experiment, we utilized the SEERA (Software enginEERing in SudAn) public dataset. It comprises data from 120 software development projects across 42 organizations in Sudan, encompassing 76 attributes covering General Information, Size, Effort, Environment, Users, Developers, Project, and Product aspects. Since 'Actual Cost' and 'Actual Duration' implicitly reflect the workload (Effort), in practical engineering scenarios, it is commonly understood that duration typically correlates monotonically with expenses. Therefore, in subsequent experiments, we exclusively employed the 'Actual Duration' attribute as the label value.

When our research is based on the data element market, it is still necessary for us to first clean the source data.  Erroneous values such as ['?', 'Not exist'] need to be replaced with null values. Subsequently, an assessment of the dataset's missing values should be conducted, and columns with a missing ratio of  $\geq$ 10\% (shown as table \ref{tab:missing_1}) should be eliminated. Additionally, rows with an excessive number of missing fields should also be removed.

\begin{table}[htbp]
  \centering
  \caption{Attributes with a missing ratio $\geq$ 10\%}.
    \begin{tabular}{|l|r|r|l|}
    \hline
    \textbf{Attribute} & \textbf{Missing Count} & \textbf{Missing Percent} & \textbf{Type} \\
    \hline
    Outsourcing impact & 108   & 90.00\% & object \\
    Estimated size & 107   & 89.17\% & float64 \\
    Degree of standards usage & 98    & 81.67\% & object \\
    \% project gain (loss) & 17    & 14.17\% & object \\
    \hline
    \end{tabular}
  \label{tab:missing_1}
\end{table}
For attributes with relatively low missing proportions ($<$10\%), we fill in the missing values according to the method outlined in the table \ref{tab:missing_2}. 

\begin{figure}
\centering
\includegraphics[width=1\linewidth]{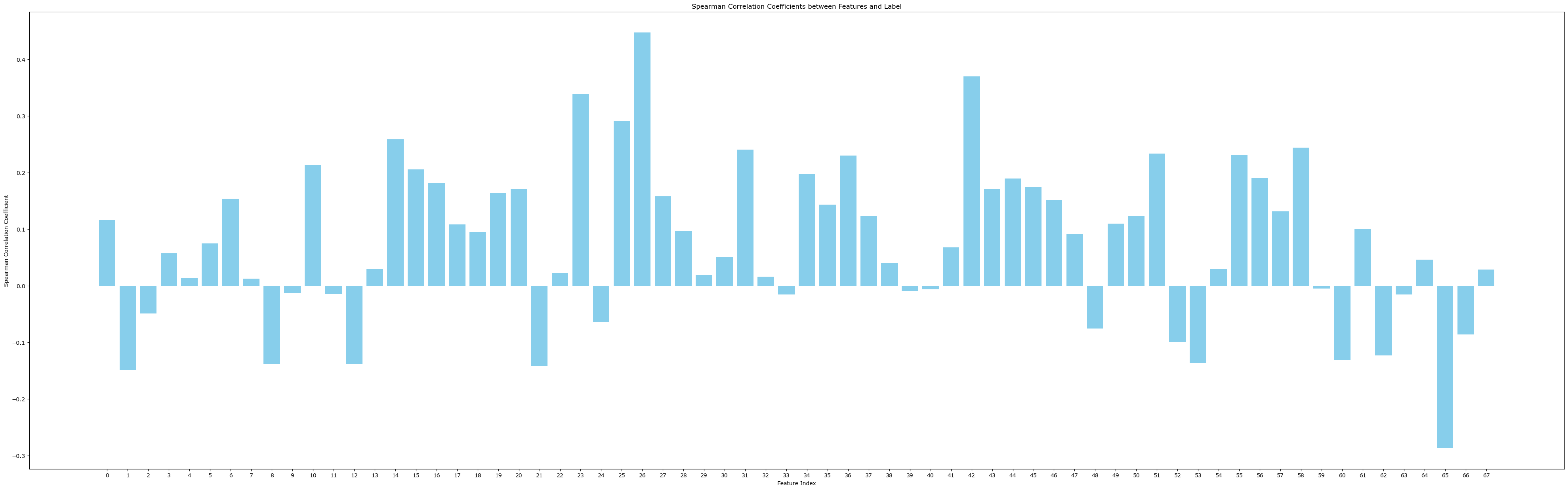}
\caption{\label{fig:s_corr}Plot of Spearman correlation analysis among SEERA dataset.}
\end{figure}

\begin{table}[htbp]
  \centering
  \caption{Imputation of missing attribute values.}
    \begin{tabular}{|l|r|r|l|}
    \hline
    \textbf{Attribute} & \textbf{Missing Count} & \textbf{Missing Percent} & \textbf{Imputation} \\
    \hline
    Team contracts & 11    & 9.24\% & Mode \\
    Requirement accuracy level & 3     & 2.52\% & Mode \\
    Process reengineering & 3     & 2.52\% & Mean \\
    Income satisfaction & 2     & 1.68\% & Mode \\
    Organization management structure clarity & 2     & 1.68\% & Mode \\
    Product complexity & 2     & 1.68\% & Mode \\
    Developer training & 2     & 1.68\% & Mode \\
    Reliability requirements & 1     & 0.84\% & Medium \\
    Comments within the code & 1     & 0.84\% & Medium \\
    Degree of software reuse & 1     & 0.84\% & Medium \\
    Team selection & 1     & 0.84\% & Medium \\
    Object points & 1     & 0.84\% & Medium \\
    Clarity of manual system & 1     & 0.84\% & Medium \\
    Development team management & 1     & 0.84\% & Medium \\
    Developer incentives policy & 1     & 0.84\% & Medium \\
    Developer hiring policy & 1     & 0.84\% & Medium \\
    Government policy impact & 1     & 0.84\% & Medium \\
    Specified H/W & 1     & 0.84\% & Medium \\
    \hline
    \end{tabular}
  \label{tab:missing_2}
\end{table}
Although the dataset contains many categorical values, since they are already represented in numerical form, there is no need for further feature encoding. The obtained dataset is now essentially usable. Before beginning experiments, we examine the Spearman correlation coefficient between the feature set and the labels. Similar to the Pearson correlation coefficient, it measures the rank correlation between two variables but does not assume a linear relationship. It first converts the variable values to ranks and then calculates the Pearson correlation coefficient between these ranks. Based on the analysis depicted in the figure \ref{fig:s_corr}, we sort the absolute values of the Spearman coefficients and obtain the top 10 features most correlated with the labels. Subsequently, we will compare the differences between feature subsets selected through various feature selection methods and this feature set. Note that we do not directly remove relatively irrelevant features such as 'ProdjId', 'Year of project', 'Open source software', etc.\cite{SEERA}, as we aim to observe whether MARLFS can automatically remove irrelevant features without introducing manual analysis.

\subsection{Evaluation Metrics }
To evaluate the performance of various algorithms or methods in the experiments, we utilize the following evaluation metrics for comparison. 

\subsubsection{MSE}
\textbf{Mean Squared Error (MSE)} is a commonly used metric for measuring the difference between the predicted and actual values in regression problems. It calculates the average of the squared errors between the predicted and actual values (shown in definition \ref{eq:mse}). A smaller MSE value indicates a smaller difference between the model's predicted results and the actual values, indicating better performance of the model. 

\subsubsection{MAE}
\textbf{Mean Absolute Error (MAE) }is another commonly used metric for measuring the accuracy of regression model predictions. It calculates the average of the absolute errors between the predicted and actual values. MAE represents the average prediction error magnitude. Unlike MSE, MAE does not consider the square of errors, making it insensitive to outliers.

The calculation formula is as follows: 
\begin{equation}
    MAE = \frac{1}{n} \sum_{i=1}^{n} |y_i - \hat{y}_i|
\end{equation}

Where $n$ is the sample size, $y_i$  is the true value of the $i$-th sample, and $\hat{y}_i$ is the predicted value of the  $i$-th sample.

\subsection{Benchmark Models }

In the experiment, we mainly compare the feature selection effects of MARLFS with Filter and Wrapper methods, as well as comparing them with expert judgment data available in the dataset. Prior to actual execution, we divide the training set and validation set in a ratio of 8:2 and standardize the feature values of both parts to expedite the model convergence speed. 

\subsubsection{Expert Judgment }
Attributes 'Estimated Cost' and 'Estimated Duration' respectively represent the estimated effort and expected duration obtained using expert judgment. In the following experiment, we will also compare our feature selection results with those obtained using the expert judgment method. 

\subsubsection{Filter }
As a comparison, for the Filter feature selection, we utilize the VarianceThreshold class from scikit-learn, which is essentially a variance selector. During the experiment application, we will set the threshold value to 1, meaning features with variance less than 1 will be filtered out. Intuitively, these features, due to their minimal variation, are considered to provide little assistance in model training. 

\subsubsection{Wrapper}
For the Wrapper method, we employ scikit-learn's RFE (Recursive Feature Elimination). It iteratively trains the model and eliminates the least important features to complete the feature selection process. The embedded regressor uses the default RandomForestRegressor implemented in scikit-learn, with 100 decision trees and Mean Squared Error (MSE) as the criterion for measuring split quality. It's important to note that the Wrapper method requires specifying the desired length of the final feature subset beforehand. Therefore, in our experiment, we compare the performance with feature subsets of lengths 10, 20, and 30 respectively. 

\subsubsection{MARLFS}
In our subsequent experiments, we employ Graph Convolutional Networks (GCN) to achieve fixed-length state representations. Similarly, the regressor utilized is the default RandomForestRegressor implemented in scikit-learn. For the Deep Q-Network (DQN), we configure a two-layer ReLU network with 64 nodes in the first layer and 8 nodes in the second layer. The Replay Buffer size is set to 2000, and each sampling from the Replay Buffer is done with a BATCH\_SIZE of 32. We use the Adam optimizer with a learning rate of 0.01, an $\epsilon$-greedy parameter of 0.9, and a discount factor of 0.9. Regarding the reward function designed in section 2.4, we set the decay coefficient $k$ to 0.01, the weight $\alpha$ to 1, and the weight $\beta$ to 0.3. Additionally, we multiply the resulting rewards by a scaling factor of 100 for better observation.

\subsection{Performance Result }
The experimental results (shown in table \ref{tab:perf_1}) are as follows: (1) 'MARLFS' represents the algorithm introduced in this experiment. (2) 'Expert Estimate' denotes the results determined by experts. (3) 'raw-RF' shows the results obtained by directly using the random forest algorithm for regression without any feature selection. (4) 'Filter' indicates the results obtained by regression using features filtered based on a variance threshold of less than 1. (5) 'Wrapper-10,20,30' respectively represent the results obtained by regression after feature selection using RFE to select 10, 20, and 30 features. 

\begin{table}[htbp]
  \centering
  \caption{Performance Comparison}
    \begin{tabular}{|l|r|r|r|r|r|r|r|}
    \hline
          & MARLFS & Expert Estimate & raw-RF & Filter & Wrapper-10 & Wrapper-20 & Wrapper-30 \\
    \hline
    MSE   & 60.70268838 & 70.01365546 & 102.8778094 & 133.1573948 & 112.9499594 & 108.0794802 & 101.8112354 \\
    MAE   & 5.915791667 & 4.903361345 & 7.615625 & 8.24979166 & 7.866041667 & 7.428958333 & 7.184166667 \\
    \hline
    \end{tabular}
  \label{tab:perf_1}
\end{table}

\section{Result Analysis }

\subsection{Comparison of Evaluation Metrics }

\begin{figure}
    \centering
    \begin{subfigure}[b]{0.4\textwidth}
        \centering
        \includegraphics[width=\textwidth]{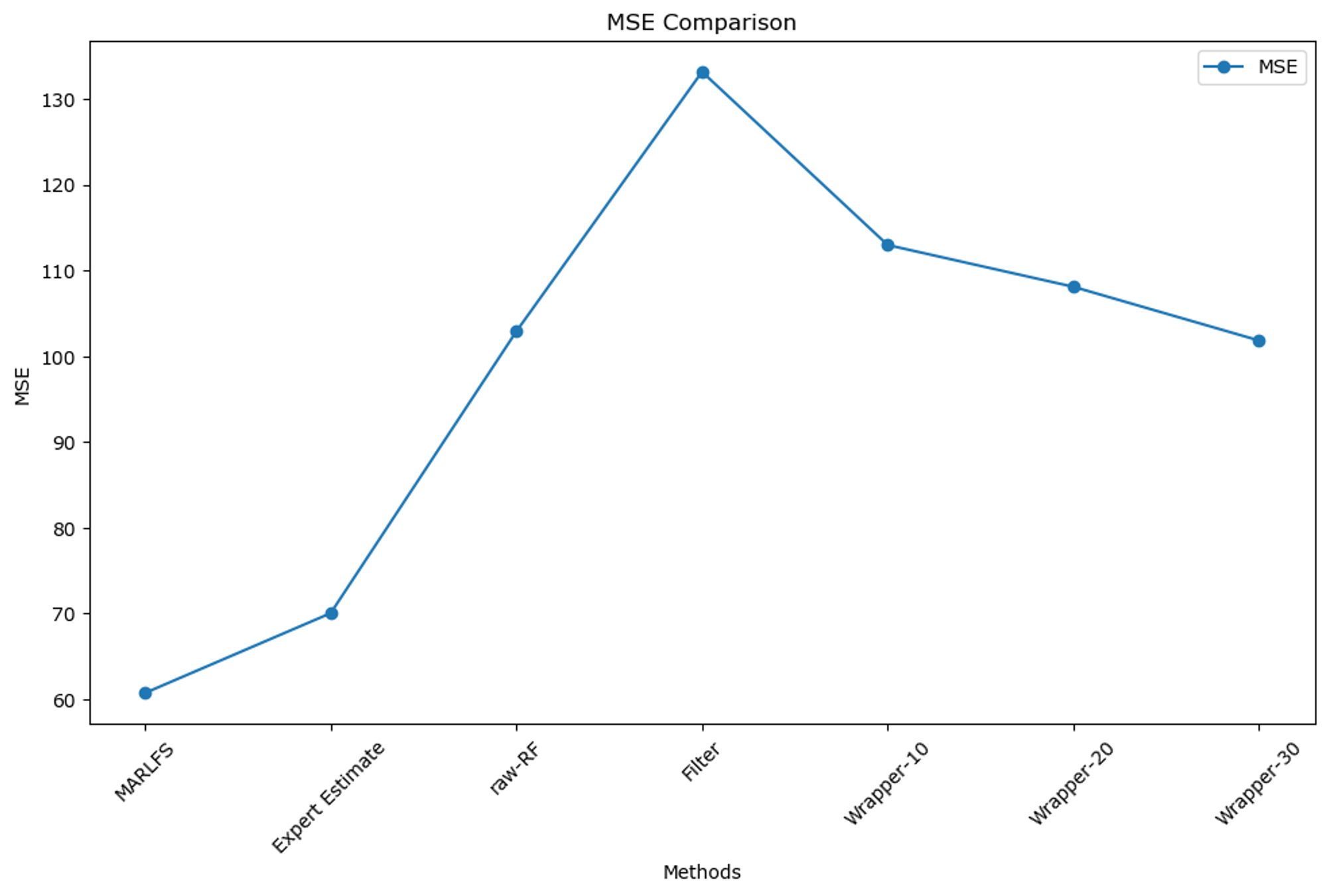}
        \caption{ methods MSE comparison }
        \label{fig:mse}
    \end{subfigure}
    \hfill
    \begin{subfigure}[b]{0.4\textwidth}
        \centering
        \includegraphics[width=\textwidth]{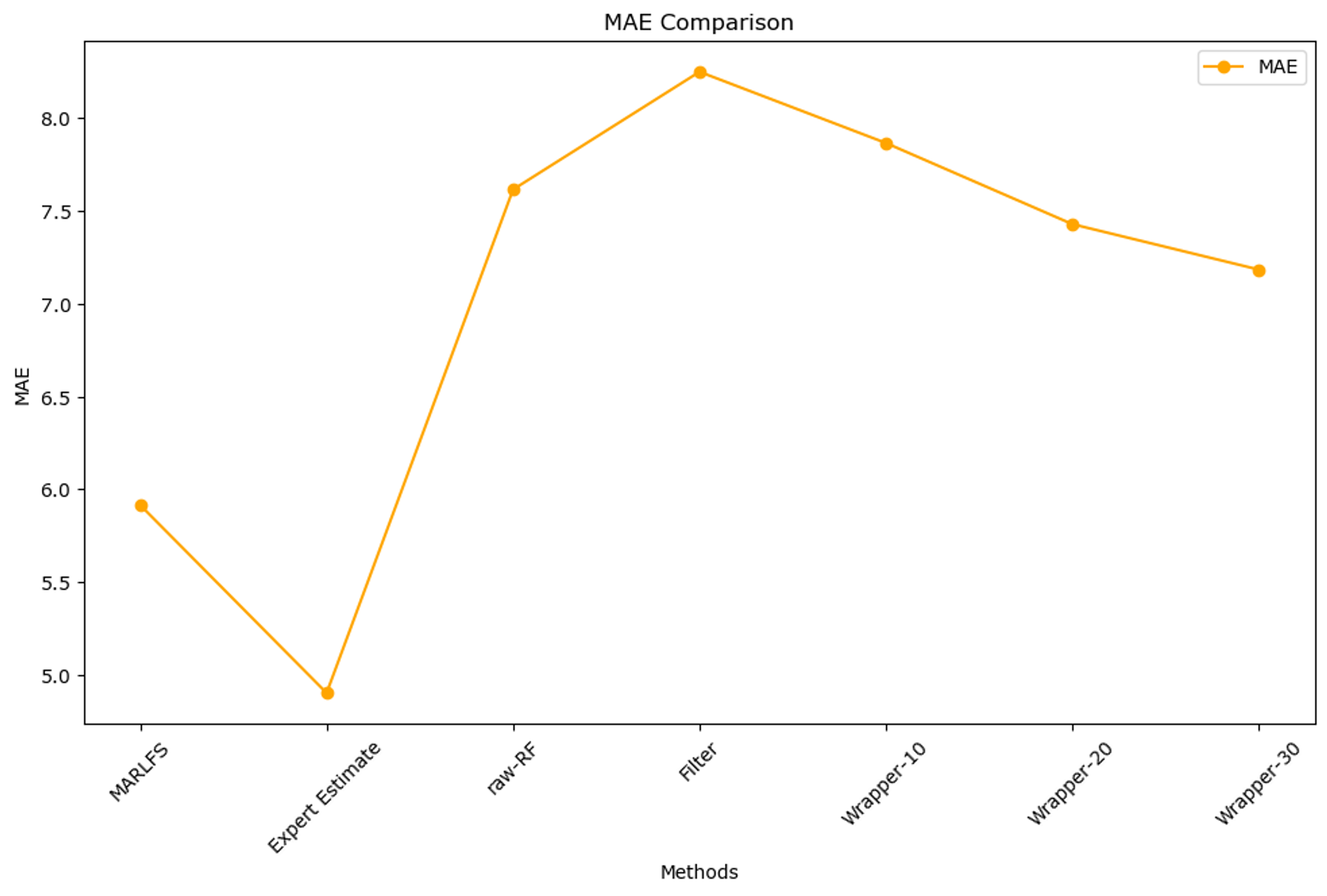}
        \caption{methods MAE comparison}
        \label{fig:mae}
    \end{subfigure}
    \caption{Comparison of experiment methods}
    \label{fig:plot_exp_com}
\end{figure}

As demonstrated in figure \ref{fig:plot_exp_com} , MARLFS achieved an MSE of 60.7 and an MAE of 5.92 on the validation set, while the MSE obtained using the expert method is approximately 70.01, with an MAE of about 4.9. Compared to the expert method, MARLFS exhibits a lower MSE but a higher MAE, indicating that MARLFS is more flexible than the expert method and can adapt to outliers, whereas the expert method tends to smooth out the overall data processing.

Without feature selection, directly training the model yielded an MSE of approximately 102.88 and an MAE of about 7.62, which is lower than the expert method but significantly higher than MARLFS. Moreover, using only the features filtered by 'Filter' resulted in even worse performance, with an MSE of only 133.16. The MSEs obtained using 'Wrapper-10', 'Wrapper-20', and 'Wrapper-30' were 112.95, 108.08, and 101.81, respectively, still lower than those obtained using expert judgment. Due to the poor performance of 'Filter', we also tested it on the training set, yielding an MSE of 13.66,  shown as table \ref{tab:perf_2}, demonstrating severe overfitting in the 'Filter' experiment.

\begin{table}[htbp]
  \centering
  \caption{Performance Comparison for Filter and Filter-train}
    \begin{tabular}{|l|r|r|}
    \hline
          & Filter & Filter-train \\
    \hline
    MSE   & 133.1573948 & 13.66145289 \\
    MAE   & 8.24979166 & 2.5013157 \\
    \hline
    \end{tabular}
  \label{tab:perf_2}
\end{table}

In summary, the experimental results suggest that employing MARLFS for software effort estimation yields prediction performance that is superior to traditional feature selection methods and more robust than expert judgment.

 \subsection{Feature Selected Analysis }
 
\begin{figure}
\centering
\includegraphics[width=1\linewidth]{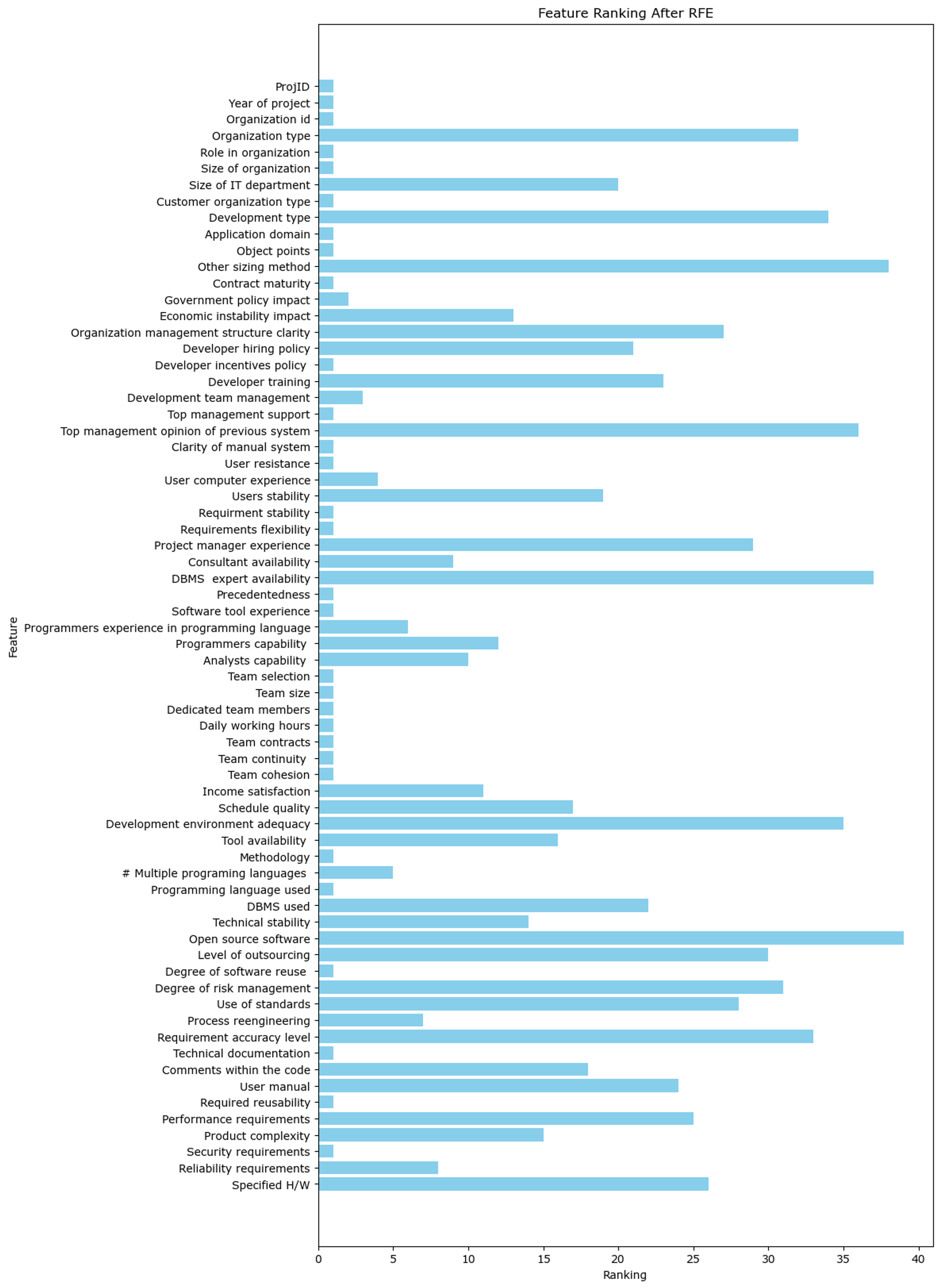}
\caption{\label{fig:RFE}Plot of feature ranking by RFE.}
\end{figure}

The comparison between the features selected by MARLFS and the top 10 features based on absolute Spearman correlation coefficients is presented in the table. It can be observed that MARLFS has filtered out 31 features from the processed feature set of 68 features, out of which 7 features match the top 10 features based on absolute Spearman correlation coefficients, as illustrated in table \ref{tab:t_1} . It is noteworthy that some irrelevant features were not selected. This demonstrates that the MARLFS feature selection method can automatically and effectively capture the relationship between features and labels, thereby achieving satisfactory predictive performance. Specifically, in the process of using RFE for wrapper feature selection, the features were also ranked in terms of importance, as shown in the figure \ref{fig:RFE}. 

\begin{table}[htbp]
  \centering
  \caption{Feature Comparison between MARLFS and Spearman Correlation Coefficient}
    \begin{tabular}{|l|l|l|}
    \hline
    Sequence & Features by MARLFS & Features by Spearman Correlation Coefficient \\
    \hline
    1 & Organization id & - \\
    2 & Size of IT department & - \\
    3 & Customer organization type & - \\
    4 & Application domain & - \\
    5 & Other sizing method & - \\
    6 & Contract maturity & - \\
    7 & Developer hiring policy & - \\
    8 & Development team management & - \\
    9 & Clarity of manual system & - \\
    10 & User resistance & User resistance \\
    11 & Requirment stability & Requirment stability \\
    12 & DBMS expert availability & - \\
    13 & Precedentedness & Precedentedness \\
    14 & Programmers capability & - \\
    15 & Team selection & - \\
    16 & Team contracts & - \\
    17 & Team cohesion & Team cohesion \\
    18 & Schedule quality & - \\
    19 & Development environment adequacy & - \\
    20 & Methodology & - \\
    21 & Programming language used & - \\
    22 & Technical stability & Technical stability \\
    23 & Level of outsourcing & - \\
    24 & Degree of risk management & Degree of risk management \\
    25 & Use of standards & - \\
    26 & Requirement accuracy level & Requirement accuracy level \\
    27 & Technical documentation & - \\
    28 & Comments within the code & - \\
    29 & Required reusability & - \\
    30 & Performance requirements & - \\
    31 & Product complexity & - \\
    32 & - & Users stability \\
    33 & - & Security requirements \\
    34 & - & Economic instability impact \\
    \hline
    \end{tabular}
  \label{tab:t_1}
\end{table}
Furthermore, by examining the 31 features selected through MARLFS, it becomes evident that these features are crucial for project management, particularly in the context of effort estimation. These features can be categorized based on their relevance to different aspects of software development and project management, as shown in the table \ref{tab:t_2} : (1) Organizational Attributes: Characteristics and size of the organization can influence project complexity and scale\cite{zuzhi1}. (2) Project Management and Team Attributes: This category includes factors related to how projects and their teams are managed. Effective team management, strong team cohesion, and mature contracts can enhance accuracy in effort estimation by reducing uncertainty and improving productivity\cite{teambuild1, teambuild2}. (3) Requirements and User Interaction: Attributes in this category involve the clarity and stability of project requirements\cite{qikan-xuqiu}, as well as user interaction. Clear and stable requirements, along with minimal user resistance, can lead to more reliable effort estimation by minimizing changes and rework\cite{xuqiu1, xuqiu2, xuqiu3}. (4) Technical Attributes: Technical attributes encompass a wide range of factors from application domain to specific technologies and standards used. These factors directly impact the complexity of development and required effort, influencing effort estimation. For example, complex products or new programming languages can increase the required effort\cite{jishu1}. (5) Outsourcing and Documentation: The level of outsourcing\cite{waibao1} and the quality of documentation (including technical documents and code comments) can significantly affect effort estimation\cite{wendang1}. High-quality documentation and strategic outsourcing can streamline the development process and reduce effort uncertainty\cite{waibao2, wendang2}. (6) Quality and Performance Requirements: These attributes focus on the non-functional requirements of the software, such as performance and reusability. Meeting high-performance or reusability standards can increase effort as it requires additional design and testing work\cite{xingneng1, xingneng2}.

\begin{table}[htbp]
  \centering
  \caption{Cluster features selected by MARLFS.}
    \begin{tabular}{|l|l|p{6cm}|}
    \hline
    Aspect Sequence & Aspect Name & Attributes \\
    \hline
    1 & Organizational Attributes & Organization id, Size of IT department \\
    2 & Project Management and Team Attributes & Contract maturity, Developer hiring policy, Development team management, Team selection, Team contracts, Team cohesion, Schedule quality, Degree of risk management \\
    3 & Requirements and User Interaction & Clarity of manual system, User resistance, Requirment stability, Requirement accuracy level, Customer organization type \\
    4 & Technical Attributes & Application domain, Other sizing method, DBMS expert availability, Precedentedness, Programmers capability, Development environment adequacy, Methodology, Programming language used, Technical stability, Product complexity \\
    5 & Outsourcing and Documentation & Level of outsourcing, Technical documentation, Comments within the code \\
    6 & Quality and Performance Requirements & Performance requirements, Required reusability, Use of standards \\
    \hline
    \end{tabular}
  \label{tab:t_2}
\end{table}

Each of these categories plays a vital role in understanding and managing software projects, thereby contributing to more accurate effort estimation through feature selection. This also highlights that the features selected through feature selection have intuitive managerial implications.

\section{Conclusion }
Through the experiments mentioned above, we can observe the potential of applying advanced feature selection algorithms to software project management, specifically in the task of software development effort estimation. Specifically, the application of feature selection to software development effort estimation holds the following significance: (1) Improved prediction accuracy: By automatically and efficiently selecting the most useful feature subsets, algorithms can assist software project managers in predicting software effort more accurately. This enables managers to have a more precise expectation during project planning, facilitating better resource allocation and schedule planning. (2) Reduced development costs: Accurate effort prediction allows better control over project development costs. Rational resource allocation and schedule planning can reduce unnecessary overtime and resource wastage, thus lowering the overall cost of project development. (3) Enhanced project delivery quality: Selecting the most useful feature subsets provides a better understanding of the key factors influencing software effort. This helps project managers to better identify and address potential risks and take corresponding measures to ensure project delivery quality. (4) Strengthening the scientific basis of management decisions: Based on precise effort prediction and identification of key features, project managers can make decisions and formulate strategies more scientifically according to the selected features. This helps project teams better respond to changes and challenges, increasing the likelihood of project success. (5) Flexible and comprehensive response to market changes: Market services in the data element market may provide new market data and trend analysis at any time. Through automated feature selection algorithms, software project teams can adjust feature subsets in a timely manner to reflect the latest market changes and respond flexibly to market competition and demand changes.

In summary, feature selection in the context of the data element market brings an innovative approach to software project management. By leveraging diverse data from the market, it enhances the accuracy and reliability of effort estimation, driving project management towards a more intelligent, data-driven direction. This approach not only improves the efficiency of project management but also provides a more scientific foundation for project decision-making.

\bibliographystyle{apalike}
\bibliography{sample}

\end{document}